\documentclass[12pt,3p]{elsarticle}
\usepackage{natbib}
\usepackage{algorithmicx}
\biboptions{sort&compress}

\usepackage{bm}
\usepackage{color}
\usepackage[normalem]{ulem}

\usepackage[lined,boxed]{algorithm2e}
\usepackage{amssymb}
\def\orion{\textsc{ORION}}

\def\hii{H~\textsc{ii}}

\def\workremains{\textsc{workremains}}
\def\alldone{\textsc{alldone}}
\def\datarecv{\textsc{datarecv}}
\def\countrecv{\textsc{countrecv}}
\def\loopiter{\textsc{loopiter}}
\def\harm{\textit{HARM${^2}$}}

\begin{document}

\title{Hybrid Adaptive Ray-Moment Method (\harm):  \\A Highly Parallel Method for Radiation Hydrodynamics on Adaptive Grids}

    \author[rvt]{A.L.~Rosen\corref{cor1}}
    \ead{alrosen@ucsc.edu}
    \ead[url]{www.anna-rosen.com}
    \author[anu]{M.R.~Krumholz}
    \author[farm]{J.S.~Oishi}
    \author[ucb]{A.T.~Lee}
    \author[ucb,llnl]{R.I.~Klein}

\address[rvt]{Department of Astronomy \& Astrophysics, University of California Santa Cruz, Santa Cruz, CA 95064, USA}
\address[anu]{Research School of Astronomy \& Astrophysics, Australian National University, Canberra, ACT 2611, Australia}
\address[farm]{Department of Physics \& Astronomy, Bates College, Lewiston, ME 04240, USA}
\address[ucb]{Department of Astronomy, University of California, Berkeley, CA 94720, USA}
\address[llnl]{Lawrence Livermore National Laboratory, PO Box L-23, Livermore, CA 94550, USA}

\begin{abstract}
We present a highly-parallel multi-frequency hybrid radiation hydrodynamics algorithm that combines a spatially-adaptive long characteristics method for the radiation field from point sources with a moment method that handles the diffuse radiation field produced by a volume-filling fluid. Our Hybrid Adaptive Ray-Moment Method (\harm) operates on patch-based adaptive grids, is compatible with asynchronous time stepping, and works with any moment method. In comparison to previous long characteristics methods, we have greatly improved the parallel performance of the adaptive long-characteristics method by developing a new completely asynchronous and non-blocking communication algorithm. As a result of this improvement, our implementation achieves near-perfect scaling up to $\mathcal{O}(10^3)$ processors on distributed memory machines. We present a series of tests to demonstrate the accuracy and performance of the method.
\end{abstract}

\maketitle
%\keywords{stars: massive stars, feedback, radiation pressure ISM: bubbles, dust, turbulence}

\section{Introduction}
Radiation-hydrodynamics (RHD) is a challenging numerical problem, but it is a crucial component in modeling several physical phenomena in the fields of astrophysics, laser physics, and plasma physics. Accurate solution of the radiative transfer (RT) equation, which governs the evolution of radiation interacting with matter, is difficult because of its high dimensionality. This equation depends on six independent variables: three spatial, two angles describing the direction of the propagation of photons, and one frequency dimension. For time-dependent RHD calculations, this solution must be obtained at every time step, and then coupled to the hydrodynamics. Even on parallel supercomputers direct solution of the RT equation at each time step of a time-dependent calculation is prohibitively expensive, because of this most numerical RHD codes use approximations to treat the evolution of the radiation field and its interaction with matter. 

One common approach to solving the RHD equations is to reduce the dimensionality of the problem. This class of approximations are known as moment methods because they take the moments of the radiative transfer equation in direct analogy to the Chapman-Enskog procedure used to derive the hydrodynamic equations from the kinetic theory of gases \citep{krumholz2011b, teyssier2015a}. This method averages over the angular dependence, and thus is a good approximation for smooth, diffuse radiation fields such as those present in optically thick media when the radiation is tightly coupled to the matter. The accuracy with which moment methods recover the angular dependence of the true solution depends on the order at which the moments are closed, and on the closure relation adopted. Common approximations include flux-limited diffusion (FLD; closure at first moment) \citep{levermore1981a,krumholz2007a,commercon11a}, the M1 method (closure at the 2nd moment using a minimum entropy closure) \citep{gonzalez2005a,rosdahl15a}, and Variable Eddington Tensor (VET; closure at the 2nd moment using an approximate solution to the full transfer equation) \citep{dykema1996a, jiang2012a, davis2012a}. Regardless of the order and closure relation, the computational cost of these methods usually scales as $N$ or $N \log{N}$, where $N$ is the number of cells, and the technique is highly parallelizable \citep{krumholz2011b}.

An alternative technique used to solve the RT equation numerically is characteristics-based ray tracing, which solves this equation directly along specific rays. With this method, the directionality of the radiative flux is highly accurate, but the accuracy depends on the sampling of rays. Two widely used schemes for ray tracing in grid-based codes are \textit{long} and \textit{hybrid} characteristics. Long characteristics traces rays on a cell by cell basis, and provides maximum possible accuracy. Hybrid characteristics is a combination of long characteristics within individual grids and short characteristics between grids (i.e., in which only neighboring grid cells are used to interpolate incoming intensities) \citep{rijkhorst2006a, buntemeyer2016a}. The method of short characteristics is faster but more diffusive compared to long characteristics methods. The computational cost for both methods scales linearly with the number of sources, rays traced, and grid cells with which the rays interact, making these methods prohibitively expensive for treating diffuse radiation fields where every computational cell is a source. Instead, they are ideal for treating the radial radiation fields of point sources. Even for this application, however, one major drawback of ray tracing methods, especially long characteristics, is that they are difficult to parallelize in a code where the hydrodynamics is parallelized by domain decomposition. In such a configuration, each ray will usually cross multiple processor domains, creating significant communications overheads and serial bottlenecks. 

In summary, moment methods are better at approximating the diffuse radiation field from a fluid but are poor at modeling the propagation of radiation from point sources where the direction of the field is important. Characteristics methods, in contrast, are good at approximating the direction-dependent radiation fields from point sources but are too computationally expensive for practical use in simulating a diffuse radiating fluid. When both point and diffuse radiation sources are present, therefore, a natural approach is to combine both techniques by using long characteristics to model the propagation of radiation from a point source and its subsequent interaction (e.g., absorption) with the fluid and then use a moment method to follow the subsequent diffuse re-emission. 

This technique has been developed in several numerical codes in the past 20 years, but these codes typically have been limited to cases where a geometric symmetry simplifies the long characteristics solution. \citet{wolfire86a, wolfire87a} introduced a formal decomposition between the direct and dust-reprocessed radiation fields for a calculation in 1D spherical geometry. The first published 2D simulation using such a method is \citet{murray1994a}, who coupled long characteristics to FLD to model the direct (ray tracer) and scattered (FLD) radiation field in accretion disk coronae. \citet{kuiper2010a} incorporated a similar hybrid approach in the 3D grid based code Pluto, but again limiting the problem to a special geometry: in this case a single point source at the origin of a spherical computational grid. Most recently, \citet{klassen2014a} developed a hybrid scheme in the FLASH adaptive mesh refinement (AMR) code but uses FLD plus hybrid characteristics which, although faster, is less accurate than long characteristic methods. 

The reason that many authors have resorted to special geometries or abandoned long characteristics is the difficulty in parallelizing long characteristics in a general geometry, particularly in the case of adaptive grids. The problem is difficult because it is unknown \textit{a priori} how far rays will travel and what grids they will interact with in an adaptive grid framework. In a distributed memory paradigm where different grids may be stored in memory on different processors, this can easily result in a complex communication pattern with numerous serial bottlenecks. Indeed, all implementations of long characteristics on adaptive grids published to date use synchronous communication algorithms in which processors must wait for other processors to receive ray information \citep{wise2011a}, leading to exactly this problem. 

In this paper we present our Hybrid Adaptive Ray-Moment Method (\harm) which uses long characteristics to treat radiation from point sources coupled to a moment method to handle the diffuse radiation field from the fluid. \harm\ works on adaptive grids with asynchronous time stepping. We have greatly improved the parallelism of the long characteristics solve in a distributed memory framework through a new, completely asynchronous, non-blocking communication method. The rest of this paper is organized as follows. We begin with a formal derivation of our method for decomposing the radiation field into two components in section \ref{sec:theory}. Section \ref{sec:num} describes our numerical implementation of our hybrid radiation scheme in the astrophysical AMR code \orion. Next we confirm the robustness of our method by providing validation and performance tests in sections \ref{sec:valid} and \ref{sec:perform}, respectively. Finally, we summarize our methods and results in section {\ref{sec:sum}}.

\section{Decomposition of the Radiation-Hydrodynamics Problem}
\label{sec:theory}
Here we describe a formal method to separate the radiation field into two components -- (1) the diffuse radiation from the fluid and (2) the direct radiation field from point sources \citep{Norman1998a}. Formally, we consider a system consisting of a volume-filling radiating fluid plus point sources of radiation, and we wish to decompose the radiation fields produced by the fluid and the point sources. An example where such a decomposition is valuable is in the problem of simulating stars embedded in an optically thick, dusty medium such as is present during the early formation of a star cluster while the stars are actively accreting. The radiative flux from the stars will be absorbed by nearby dust and the dust will re-emit thermal radiation in the infrared. This radiation will be highly coupled to the interstellar medium and diffuse through the dense gas.

We begin with the equations of radiation-hydrodynamics (RHD) written in the lab-frame \citep{mihalas1982a, mihalas2001a}:
\begin{eqnarray}
\label{eqn:mass}
\frac{\partial \rho}{\partial t} + \nabla \cdot \left(\rho \mathbf{v} \right) &=& 0 \\
\label{eqn:mom}
\frac{\partial}{\partial t} \left( \rho \mathbf{v}\right) + \nabla \cdot \left( \rho \mathbf{vv} \right) &=& -\nabla P + \mathbf{G} \\
\label{eqn:ene}
\frac{\partial}{\partial t} \left( \rho E \right) + \nabla \cdot  \left[  \left(\rho E + P  \mathbf{v} \right) \right] &=&  cG^0
\end{eqnarray}
\noindent
where $\rho$, $\mathbf{v}$, $E$, and $P$ are the density, velocity, specific energy (thermal plus kinetic), and thermal pressure of the fluid, respectively; and $\left(G^0, \mathbf{G} \right)$ is the radiation four-force density which is the negative of the radiation energy stress tensor and is given by
\begin{eqnarray}
\label{eqn:4force1}
cG^0 &=& \int^\infty_0 d\nu \int d\Omega \left[ \kappa(\mathbf{n}, \nu ) I(\mathbf{n}, \nu ) - \eta (\mathbf{n}, \nu )\right] \\
\label{eqn:4force2}
c\mathbf{G} &=& \int^\infty_0 d\nu \int d\Omega \left[ \kappa(\mathbf{n}, \nu ) I(\mathbf{n}, \nu ) - \eta (\mathbf{n}, \nu )\right] \mathbf{n}
\end{eqnarray}
\noindent
where $I(\mathbf{n}, \nu )$ is the intensity of the radiation field at frequency $\nu$ in direction $\mathbf{n}$. We note that the physical quantities given in Equations (\ref{eqn:mass})-(\ref{eqn:ene}) depend on spatial position and time.
The time-like and space-like components of $\left(G^0, \mathbf{G} \right)$ represent the rate of energy and momentum transfer from the radiation to the fluid, respectively. The intensity is governed by the time-independent radiative transfer equation
\begin{equation}
\label{eqn:rad}
\textbf{n} \nabla I(\nu,\mathbf{n}) = -\kappa(\mathbf{n}, \nu ) I(\mathbf{n}, \nu ) + \eta (\mathbf{n}, \nu )
\end{equation}
\noindent 
where $\kappa(\mathbf{n}, \nu )$ and $\eta(\mathbf{n}, \nu )$ are the direction and frequency dependent absorption and emission coefficients in the lab-frame, respectively. For simplicity, we have neglected the effects of scattering because we expect to solve the equations of RHD in astrophysical problems where absorption is the dominant transfer mechanism. However, it would be straightforward to extend the method to include scattering in the diffuse component, as we point out below. We also ignore the time-dependence of the radiative transfer equation because our primary target application is systems where the light travel time is orders of magnitude smaller than the system dynamical time, and thus the radiation intensity is always in instantaneous equilibrium. 

We now separate $I(\mathbf{n}, \nu)$ into two components
\begin{equation}
I(\mathbf{n}, \nu) = I_{\rm dir}(\mathbf{n}, \nu) + I_{\rm diff}(\mathbf{n}, \nu)
\end{equation}
\noindent 
to describe the direct radiation fields from point sources ($I_{\rm dir}$) and the diffuse radiation field ($I_{\rm diff}$) emitted by the fluid. Since the sources that contribute to the direct radiation field are point sources we can represent their intensity as a sum of $\delta-$functions
\begin{equation}
I_{\rm dir}(\mathbf{n},\nu) = \sum^N_{i=1} I_{\mathrm{src},i}(\nu) \delta\left(\mathbf{n} - \mathbf{n}_{\mathrm{src},i}\right),
\end{equation}
where $\mathbf{n}_{\mathrm{src},i} = (\mathbf{x} - \mathbf{x}_i) / | \mathbf{x} - \mathbf{x}_i |$ for any position $\mathbf{x}$ in the computational domain, $\mathbf{x}_i$ and $I_{\mathrm{src},i}$ are the position and intensity of the $i$th point source, and we assume that the sources are isotropic emitters, so $I_{\mathrm{src},i}$ is independent of $\mathbf{n}.$\footnote{Note that this limits our method to non-relativistic problems, where we can neglect the effects of relativistic beaming of the source radiation fields.} With this formulation $I_{\rm dir}$ is non-zero only at special values of $\mathbf{n}$, such as along radial directions between the point sources and position $\mathbf{x}$, and zero for all others; while $I_{\rm diff}$ will be non-zero everywhere. However, because the four-force vector $(G^0, \mathbf{G})$ depends on integrals over $\mathbf{n}$, the $\delta$-function contributions from $I_{\rm dir}$ may dominate at some positions, while the contribution from $I_{\rm diff}$ dominates elsewhere. This makes solution with a pure moment method difficult, and motivates us to treat the radiation fields of the point sources and fluid separately so that we can properly take into account the direction of the radiation fields from point sources. With this decomposition Equations (\ref{eqn:4force1})-(\ref{eqn:4force2}) become
\begin{eqnarray}
\label{eqn:4force11}
cG^0 &=& \int^\infty_0 d\nu \int d\Omega \left[\kappa(\mathbf{n}, \nu ) I_{\rm dir}(\mathbf{n}, \nu) - \eta_{\rm dir} (\mathbf{n}, \nu )\right]   \nonumber \\
	          && + \int^\infty_0 d\nu \int d\Omega \left[ \kappa(\mathbf{n}, \nu ) I_{\rm diff}(\mathbf{n}, \nu ) - \eta_{\rm diff} (\mathbf{n}, \nu )\right] \\
\label{eqn:4force22}
c\mathbf{G} &=& \int^\infty_0 d\nu \int d\Omega \left[\kappa(\mathbf{n},  \nu ) I_{\rm dir}(\mathbf{n}, \nu) - \eta_{\rm dir} (\mathbf{n}, \nu )\right] \mathbf{n}  \nonumber \\
	          && + \int^\infty_0 d\nu \int d\Omega \left[ \kappa(\mathbf{n}, \nu ) I_{\rm diff}(\mathbf{n}, \nu ) - \eta_{\rm diff} (\mathbf{n}, \nu )\right] \mathbf{n}
\end{eqnarray}
\noindent
where $\eta_{\rm dir}(\mathbf{n}, \nu)$ and $\eta_{\rm diff}(\mathbf{n}, \nu)$ describes the emission due to point sources and the fluid, respectively.

This decomposition allows the following general approach to a hybrid scheme: (1) use a long characteristics method to solve for $I_{\rm dir}$, (2) use a moment method to solve for $I_{\rm diff}$, (3) add the two components to obtain the radiation four-force density $(G^0, \mathbf{G})$, (4) update the hydrodynamic state using the radiation four-force density. As a further benefit to this approach, we note that there is no requirement that steps (1) and (2) use the same frequency resolution, since $(G^0, \mathbf{G})$ depends only on an integral over frequency. It is relatively straightforward to bin the intensity from the point sources by frequency with a ray tracer since each ray can be approximated by an array of intensities, while using a lower frequency resolution in the (generally more expensive) moment method. This is ideal for point sources such as stars which have color temperatures much higher than the absorbing medium. 

\section{The \harm\ Algorithm}
\label{sec:num}
In this section we describe the \harm\ algorithm. We have implemented this algorithm in the \orion\ astrophysical adaptive mesh refinement (AMR) code \citep{Klein1999a, fisher02a, krumholz2007a, li12a} and we use this implementation for all the algorithm tests described below. \orion\ uses grid-based adaptivity \citep{berger84a, berger89a} with individual time steps for each level, and the \harm\ algorithm can be applied to any AMR code following this design. Variable definitions from this section are defined in table \ref{tab:def}. \orion\ uses the FLD approximation for its moment method \citep{krumholz2007a}, and we will use this for all tests below, but \harm\ is equally compatible with any other moment method.

\subsection{Update Cycle}
Consider an adaptive mesh covering some computational domain of interest. The mesh consists of levels with different cells sizes, with $l=0$ denoting the coarsest level and $l=l_{\rm max}$ the finest. Each level, in turn, is made up of a union of rectangular grids, each with the same cell size. In a distributed-memory parallel computation, different grids may be stored in memory on different processors or nodes. The grids on a given level need not be contiguous, but they are required to be non-overlapping, and the grids are properly nested such that a cell of a level $l$ grid may have as its neighbor the domain edge, another level $l$ cell, or a cell of level $l-1$ or $l+1$, but not a cell of any other level. Point sources are only placed at locations covered by a grid of level $l_{\rm max}$. Each level advances on a time step $dt_l$, ordered such that $dt_l \geq dt_{l+1}$, and so that, after some number of time steps on level $l+1$, the simulation time $t_{l+1}$ on that level will be equal to the time $t_l$ on the next coarsest level. That is, we require that, a level $l+1$ syncs up in time with the next coarsest level $l$. In all the tests we perform with \orion\ the time steps obey $dt_{l+1} = dt_l/2$, and synchronization occurs every 2 fine time steps, but this is not required by \harm.
\\

Given this setup, our algorithm is as follows:
\noindent
\begin{enumerate}
	\item Operator split the direct and diffuse components of the radiation field:
	\begin{enumerate}
	\item if $l$ equals 0 \textbf{or} $t_{\rm start, \it l}$ is greater than $t_{\rm start, \it l \rm{-1}}$, where $t_{\rm start, \it i}$ is the current time on level $i$, then
	\begin{enumerate}
	\item Loop over point sources and inject rays onto grids that belong to level $l_{\rm max}$ where they are located.
	\item Advance rays across grids on level $l_{\rm max}$ and all coarser grids that the rays interact with, store the rates at which radiative energy and momentum, $dE/dt$ and $d\mathbf{p}/dt$, are absorbed by the gas (Section \ref{sec:art}).
	\item Restrict $dE/dt$ and $d\mathbf{p}/dt$ from finer level grids down to level $l$.
	\end{enumerate}
	\item Add $(dE/dt) \, dt_l$ and $(d\mathbf{p}/dt)\,dt_l$ to the gas energy and momenta, respectively.
	\item Update the diffuse radiation field with a moment method.
	\end{enumerate}
	\item Apply hydrodynamics update to all cells on level $l$.
	\item Update point sources if $l=l_{\rm max}$.
\end{enumerate}

\noindent
For the pattern of time steps used by \orion, whereby there are 2 fine time steps per coarse time step, this method results in $2^{l_{\rm max}}$ ray trace updates per update on the coarsest level. Note that, because we only perform a ray trace if $t_{\rm start, \it l}>t_{\rm start, \it l \rm{-1}}$, we do not perform any redundant ray tracing steps. In other words, we perform the ray trace at a given time only if we have not already performed it at that time.
 
\subsection{Direct Radiation Field: Adaptive Ray Trace}
\label{sec:art}
We now describe the adaptive ray tracing procedure that forms step 1a(ii) of the algorithm above. Consider a single point source with a specific luminosity  given by $L_\nu$ and luminosity given by $L_{\rm \star}=\int^{\infty}_{0} L_\nu \, d\nu$. The generalization to multiple point sources is trivial. We discretize the point source spectrum in frequency into $N_\nu$ frequency bins, with the $i$th bin covering a range in frequency $(\nu_{i-1/2}, \nu_{i+1/2})$. Let $L_i = \int_{\nu_{i-1/2}}^{\nu_{i+1/2}} L_\nu \, d\nu$ be the luminosity of the point source integrated over the $i$th frequency such that $\sum L_i=L_{\rm \star}$. We generally expect that $L_i$ will be the energy radiated per unit time in a given frequency bin, but the algorithm is identical if we instead take $L_i$ to be a photon luminosity, measured in photons per unit time.

We wish to solve the transfer equation along rays that end at this source. Along a ray characterized by a direction $\mathbf{n}$ and a solid angle $\Omega_{\rm ray}$ that it subtends, the propagation of the radiation is described by the time-independent transfer equation (i.e., Equation (\ref{eqn:rad})), with the emission term $\eta$ set to zero because we are taking the direct radiation field to have zero emissivity except at the point sources. Multiplying both sides of this equation by $4\pi r^2/\Omega_{\rm ray}$, we obtain an integrated form of the transfer equation
\begin{equation}
\label{eqn:rtray}
\frac{\partial L_{{\rm ray},i}}{\partial r} = -\kappa_i L_{{\rm ray},i},
\end{equation}
where $L_{{\rm ray},i}(r)$ is the luminosity along the ray at a distance $r$ from the point source and $\kappa_i$ is the total absorption opacity for the $i$th frequency bin in units of $\rm{cm}^{-1}$. This equation is subject to the boundary condition $L_{{\rm ray},i}(0) = L_i/N_{\rm pix}$, where $N_{\rm pix} = 4\pi/\Omega_{\rm ray}$. We solve this equation by discretizing it along the line segments defined by the intersection of the ray with the cells of the computational mesh. Specifically, when a ray with luminosity $L_{{\rm ray},i}$ passes through a cell along a segment of length $dl$, the optical depth of the segment is $\tau_i = \kappa_i \, dl$, and the luminosity of the ray decreases by an amount
\begin{equation}
dL_{{\rm ray},i} = L_{{\rm ray},i} \left(1 - e^{-\tau_i}\right).
\end{equation}
In the process, the cell absorbs an amount of energy and momentum at a rate
\begin{eqnarray}
\frac{dE}{dt} & = & \sum_{i=1}^{N_\nu} dL_{{\rm ray},i} \\
\frac{d\mathbf{p}}{dt} & = &  \sum_{i=1}^{N_\nu} \frac{dL_{{\rm ray},i}}{c} \mathbf{n}.
\end{eqnarray}
The total absorption rate for each cell is simply the sum of $dE/dt$ and $d\mathbf{p}/dt$ over all rays from all point sources that pass through it. When computing the line segments $dl$, we only consider grids that are not masked by any finer grid. That is, when solving Equation (\ref{eqn:rtray}), we only ever consider the most highly spatially resolved data at any given position.

We choose the directions $\mathbf{n}$ and solid angles $\Omega_{\rm ray}$ using the angular discretization introduced by \citet{abel2002a} and \citet{wise2011a}. In this approach, $\mathbf{n}$ and $\Omega_{\rm ray}$ are chosen using the Hierarchical Equal Area isoLatitude Pixelization of the sphere (HEALPix) scheme \citep{gorski2005a}, which divides the surface area of a sphere into equal area pixels that can be further subdivided into four equal-area sub-pixels. There are $N_{\rm pix}(0) = 12$ pixels at the coarsest HEALPix level, and there are $N_{\rm pix}(j) = 12\times 4^j$ pixels on HEALPix level $j$; note that the HEALPix level $j$ and the AMR grid level $l$ are distinct and in general are not the same. The scheme is adaptive in that, as we trace rays away from point sources, we subdivide them as needed to ensure that cells are adequately resolved. Specifically, we divide a ray into 4 sub-rays if it satisfies the condition
\begin{equation}
\label{eqn:split}
\frac{\Omega_{\rm cell}}{\Omega_{\rm ray}} = \frac{N_{\rm pix}(j)}{4\pi} \left(\frac{\Delta x }{r}\right)^2 < \Phi_{\rm c},
\end{equation}
\noindent
where $\Omega_{\rm cell}= (\Delta x/r)^2$ is the solid angle subtended by a cell of linear size $\Delta x$ at a distance $r$ from the point source. The quantity $\Phi_{\rm c}$ is the minimum number of rays required to go through each cell, which we usually set to $4$ following the resolution tests of \citet{krumholz2007c} and \citet{wise2011a} but, in general, the exact value for $\Phi_{\rm c}$ is problem-dependent. The initial luminosity per ray for frequency bin $i$ is $L_{{\rm ray},i,j_0} = L_{i}/N_{\rm pix}(j_0)$ where $j_0$ is the initial healpix level. When a ray splits, we solve the transfer equation along the sub-rays using the boundary condition $L_{{\rm ray},i,j+1}(R) = L_{{\rm ray},i,j}(R)/ 4$, where $L_{{\rm ray},i,j}(R)$ is the luminosity of the ray at frequency bin $i$ on HEALPix level $j$. As proposed by \citet{krumholz2007c}, we randomly rotate the orientation of the rays every time they are cast to minimize errors due to discretization in angle. Finally, we terminate the ray trace when either a ray exits the computational domain, or when $L_{{\rm ray},j}(r) < 0.001 L_{{\rm ray},j}(0)$ where $L_{{\rm ray},j}(0) = \sum_i L_{i}/\left(12 \times 4^{j-j_0}\right)$, i.e., when 99.9\% of the energy originally assigned to that ray on ray level $j$ has been absorbed.
\subsection{Parallelization}
\label{sec:parallel}

Thus far the algorithm we have described is substantially identical to that of \citet{wise2011a}. However, we adopt a very different, and much more efficient strategy to parallelize this procedure. The primary challenge to parallelizing this algorithm is avoiding serial bottlenecks. The grids through which the rays must be traced may be distributed across any number of processors, and solution of Equation (\ref{eqn:rtray}) is an intrinsically serial process because the rate of change of the energy and momentum in any cell due to radiation arriving along a particular ray depends upon the properties of all cells that lie between the point source and the cell in question. Since the numbers and positions of point sources and computational grids in the AMR structure, and their distribution in memory, are not known \textit{a priori}, minimizing bottlenecks requires an opportunistic approach: rays should be able to be processed in arbitrary order, with each processor performing ray tracing given the data available to it, and waiting for other processors only when no useful work can be done. To this end, communication must be non-blocking and asynchronous. At the same time, each process must be able to determine when the entire ray trace has been completed, so that it can proceed to the remainder of the update cycle (the moment method, hydrodynamics, etc.), and this determination must be robust against race conditions.
 
Recall that we consider the tracing of a particular ray done when it either exits the computational domain or when 99.9\% of its energy has been absorbed. To handle the problem of determining when the algorithm should terminate without relying on blocking communication, we \textit{pretend} we know how many rays \textit{could} be created by computing a maximum number of rays to be used as a counter:
\begin{equation}
\label{eqn:nmax}
N_{\rm max} = N_{\rm src} \times 12 \times 4^{j_{\rm max}}
\end{equation}
\noindent 
where $N_{\rm src}$ is the number of sources and $j_{\rm max}$ is the maximum HEALPix level we allow; we set this to 20 in all of our tests. Our algorithm involves accounting for ``all rays" that are destroyed on each processor by computing
\begin{equation}
\label{eqn:ndes}
N_{\rm destroyed, \mathit{k}} = \sum_{N_{\rm ray, \mathit{k}}} 4^{j_{\rm max} - j}
\end{equation}
\noindent 
where $k$ is the processor number, $j$ is the HEALPix level of the ray that is deleted due to absorption or leaving the computational domain, and $N_{\rm ray, \mathit{k}}$ is the total number of rays that have been deleted on processor $k$. This information is communicated to all other processors. Once the total destroyed, 
\begin{equation}
N_{\rm destroyed} = \sum_{N_{\rm CPU}} N_{\rm destroyed, \mathit{k}}, 
\end{equation}
on each processor equals $N_{\rm max}$ the ray trace is complete.

With this bookkeeping method understood, we present our message passing scheme as algorithm \ref{alg:parallel}. Note that this algorithm requires version 3.0 or later of the Message Passing Interface (MPI) standard, because only that version supports some of the non-blocking communications we require (e.g., \textbf{MPI\_Iprobe}; see Algorithm \ref{alg:parallel}). A detailed description of our parallelization strategy follows. Every processor has 4 flags that can change: (i) \alldone\ which is initially set to \textbf{false} and will be set to \textbf{true} once all rays have been destroyed, (ii) \workremains\ which is set to \textbf{true} if rays exist on this processor and \textbf{false} otherwise, (iii) \datarecv\ which is set to \textbf{true} when the processor can receive rays from other processor(s), and (iv) \countrecv\ which is set to \textbf{true} when the processor can receive $N_{\rm destroyed, \mathit{k}}$, the number of rays destroyed from other processor(s). Every processor also has a counter, \loopiter, which tracks the number of times the parallelization algorithm is iterated over.

\bigskip
\noindent
For each processor, our algorithm is as follows:
\begin{enumerate}
	\item Inject rays from point sources to grids.
	\item Compute $N_{\rm max}$ (Equation (\ref{eqn:nmax})). 
	\item Set \alldone\ to \textbf{false},  set \loopiter\ = 0, and enter outer \textbf{while}(not \alldone) loop.
	\begin{enumerate}
		\item Set \workremains\ to \textbf{true} if rays exist on grids that belong to this processor, otherwise set \workremains\ to \textbf{false}.
		\item If \workremains\ is \textbf{true} enter \textbf{while}(\workremains) loop.
		\begin{enumerate}
		\item Loop over all grids that belong to this processor and \textbf{for each} grid advance all rays that belong to that grid until they (i) leave the grid and need to be moved to another grid, (ii) become extinct, or (iii) leave the domain. Every time (ii) or (iii) occurs for a ray we delete the ray and increase the value of $N_{\rm destroyed, \mathit{k}}$ on this processor per Equation (\ref{eqn:ndes}). If (i) occurs we determine the new grid and processor for that ray.
		\item Set \workremains\ to \textbf{false}.
		\item Loop over rays that must be transferred to other grids and check if they belong to a grid on this processor or another processor. If the former is true place the ray on the correct grid and set \workremains\ to \textbf{true}, otherwise the ray must be transferred to another processor so we place the ray in a linked list, \textbf{outgoing\_ray\_list},  to be communicated to the other processors.
		\item \textbf{Repeat} until all rays have been either destroyed and/or placed into \textbf{outgoing\_ray\_list}.
		\end{enumerate}
		\item Loop over rays in \textbf{outgoing\_ray\_list}. Place rays from \textbf{outgoing\_ray\_list} into a contiguous array for MPI communication (one array per receiving processor) and perform a non-blocking \textbf{MPI\_Isend} for each array to send it to the appropriate processor.
		\item Set \datarecv\ to \textbf{true} and  if (\loopiter\ \textbf{modulo} $N_{\rm CPU}$ == 0)\footnote{This requirement reduces the time spent on performing MPI\_Iprobes which becomes more expensive as $N_{\rm procs}$ increases.} then enter the \textbf{while}(\datarecv) loop to begin receiving rays from other processors.
		\begin{enumerate}
		\item Probe other processors with the non-blocking  \textbf{MPI\_Iprobe} function to see if rays need to be received, if this is true set \datarecv\ to \textbf{true} otherwise set it to \textbf{false}.
		\item If \datarecv\ is \textbf{true} then receive rays on this processor with \textbf{MPI\_Recv}.  Place the incoming rays onto the new grids they belong to.
		\item \textbf{Repeat} until \textbf{MPI\_Iprobe} returns false.
		\end{enumerate}
	\item If this processor sent rays to another processor then perform non-blocking \textbf{MPI\_Testsome} to test for some given ray send requests to complete.
	\item If \workremains\ is false then perform non-blocking \textbf{MPI\_Testsome} to test for some of the $N_{\rm destroyed, \;k}$ send requests to complete.
	\item Set \countrecv\ to \textbf{true} and  if (\loopiter\ \textbf{modulo} $N_{\rm CPU}$ == 0) then enter the \textbf{while}(\countrecv) loop to begin receiving rays from other processors.
	\begin{enumerate}
		\item Probe other processors with the non-blocking  \textbf{MPI\_Iprobe} function to see if the value of $N_{\rm destroyed, \mathit{k}}$ on those processors needs to be received. If this is true set \countrecv\ to \textbf{true}, otherwise set it to \textbf{false}.
		\item Receive $N_{\rm destroyed, \mathit{k}}$ from processor $k$ on this processor with \textbf{MPI\_Recv} if \countrecv\ is \textbf{true} and assign to the $k$th element of an array \textbf{rayDestProc} containing $N_{\rm CPU}$ elements.
		\item \textbf{Repeat} until \textbf{MPI\_Iprobe} returns false.
	\end{enumerate}
	\item If $N_{\rm destroyed, \mathit{k}}$ on this processor has increased in the outer loop iteration then send this number with a non-blocking \textbf{MPI\_Isend} to all other processors.
	\item Compute sum of  \textbf{rayDestProc}. If this value equals $N_{\rm max}$ terminate the outer loop, else repeat outer loop and increment \loopiter\ by 1.
	\end{enumerate}
\end{enumerate}

\begin{algorithm}
\label{alg:parallel}
\SetKwRepeat{Do}{do}{while}
\SetKwData{Left}{left}\SetKwData{This}{this}\SetKwData{Up}{up}
\SetKwFunction{Union}{Union}\SetKwFunction{FindCompress}{FindCompress}
\SetKwInOut{Input}{input}\SetKwInOut{Output}{output}
\KwData{Rays and destroyed ray counts}
\BlankLine
{compute maxRays}\;
all\_done = False\;
\Do{not all\_done}{
	\Do{work\_remains}{
		\ForEach{grid that belongs to this processor} {
			advance all rays, return number ``destroyed" and add to destroyedCount\;
		} %end foreach
		check grids to see if work\_remains;
	} %end while wR
	Non-blocking MPI\_Isend rays to other processors\;
	\Do{MPI\_Iprobe  for rays returns true}{
		Non-blocking MPI\_Iprobe other processors for rays\;
		\If{MPI\_Iprobe returns true} {
			Blocking MPI\_Recv(rays)\;
		}
	}
	Non-blocking MPI\_Testsome rays MPI\_Isend requests\;
	Non-blocking MPI\_Testsome destroyed counts MPI\_Isend requests\;
	\Do{MPI\_Iprobe for destroyed counts returns true}{
		Non-blocking MPI\_Iprobe other processors for destroyed counts\;
		\If{MPI\_Iprobe returns true}{
			Blocking MPI\_Recv(processor destroyed counts)\;
		}
	}
	\If{destroyedCount greater than previous destroyedCount}{
		Non-blocking MPI\_Isend(destroyedCount to all processors)\;
	}
 	\If{no work\_Remains and Ray Send Requests == 0 and Destroyed Count Send Requests == 0  and sum(destroyedCount) == maxRays}{
		all\_done = True\;
	}
}
\caption{Asynchronous parallelization algorithm developed for the communication of rays to other processors.}
\end{algorithm}

\begin{table*}[t!]
	\begin{center}
	\caption{
	\label{tab:def}
Variable definitions used in section \ref{sec:num}
	}
	\begin{tabular}{ l l}
	\hline	
	Variable & Description \\
	\hline
	\hline
	$dl$ & Path length of ray across cell\\
	$dE/dt$ & Energy absorbed by fluid in cell from direct radiation \\
	$d\mathbf{p}/dt$ & Momenta absorbed by fluid in cell from direct radiation\\
	$dL_{\rm ray, \it{i}}$ & Absorbed luminosity from ray for the $i$th frequency bin\\
	$dt_i$ & Time step on level $i$\\
	$j$ & HEALPix ray level \\
	$j_0$ & Initial HEALPix ray level \\
	$j_{\rm max}$ & ``Maximum" ray level for adaptive ray trace\\
	$\kappa_{i}$ & total absorption opacity for the $i$th  frequency bin\\
	$l$ & AMR Level\\
	$l_{\rm max}$ & Maximum AMR level\\
	$L_{\rm \nu}$ & Specific luminosity of point source\\
	$L_{i}$  & Luminosity of point source integrated over $i$th frequency bin\\
	$L_{ray,i}$  & Luminosity in $i$th frequency bin along ray\\
	$\mathbf{n}$ & Normal direction of ray\\
	$N_{\rm CPU}$ & Number of processors \\
	$N_{\rm destroyed, \it{k}}$ & Number of rays ``destroyed" on processor $k$\\
	$N_{\rm destroyed}$ & Number of rays ``destroyed" on all processors \\
	$N_{\rm max}$ & Maximum number of rays used as a counter\\
	$N_{\rm \nu}$ & Number of frequency bins used for adaptive ray trace \\
	$N_{\rm pix}(j)$ & Number of HEALPix pixels on level $j$\\
	$N_{\rm ray, \it{k}}$ & Number of rays deleted on processor $k$\\
	$N_{\rm src}$ & Number of point sources\\
	$\Delta x$ & linear size of cell\\
	$\Phi_{\rm c}$ & Angular resolution of ray trace [rays/cell]\\
	$\Omega_{\rm cell}$ & Solid Angle subtended by a cell\\
	$\Omega_{\rm ray}$ & Solid angle associated with ray\\
	$\tau$ & Optical depth of cell\\
	\hline
	\end{tabular}
	\end{center}
%\tablecomments{
%Comments
%}
\end{table*}

\section{Validation Tests}
\label{sec:valid}
In this section we demonstrate the accuracy of our adaptive ray tracing algorithm and the absorption of the direct radiation field from point sources by performing two tests. In the first test, we turn off absorption of the radiation field to trace the radiative flux from a point source located at the center to demonstrate that our method recovers the correct $r^{-2}$ fall-off of the flux. The second test focuses on the coupling of the hydrodynamics with the adaptive ray trace. We set $\Phi_{\rm c}=4$ for all validation tests in this section, we use only a single frequency bin, and we disable the moment method, focusing only on the ray-tracing part of the algorithm. Tests for the moment method have been presented in \citet{krumholz2007a}.

\subsection{Flux Test}
\label{sec:flux}
To demonstrate the accuracy of our adaptive ray trace and its ability to maintain spherical symmetry, we place a point source of luminosity $L_{\rm \star}$ at the origin of a cubical domain extending from $-1$ to $+1$ pc in all directions. We set the opacity in all cells to zero and take $L_{\rm \star}=10^6 \; L_{\rm \odot}$ where $L_{\rm \odot}=3.84 \times 10^{33} \rm{erg \; s^{-1}}$ is the luminosity of the Sun. We use a base resolution of $128^3$ and two levels of refinement. We refine cells that are located within 16 cells from the star. 

In this setup, the total energy contained in the region that is a distance $<r$ from the origin should be exactly $E_{\rm exact}(<r) = L_{\rm \star} r / c$, where the quantity $r/c$ is simply the light-crossing time of the distance $r$. We can compare this to the total energy in this region returned by our code, which we can compute by noting that the radiation energy density of a given cell that is traversed by a series of rays is
\begin{equation}
\label{eqn:ecell}
U_{\rm rad} = \sum_{i} \frac{L_{\rm ray,\it{i}}}{dV}\frac{dl_i}{c}
\end{equation}
\noindent
where $L_{\rm ray, \it{i}}$ is the luminosity for ray $i$, $dV$ is the cell volume, and $dl_i$ is the path length of the ray through the cell. The total energy contained within cells whose distance $r_j$ from the origin is $<r$ is then
\begin{equation}
E_{\rm num}(<r) = \sum_{{\rm cells},\, r_j < r} U_{{\rm rad},j}\, dV
\end{equation}
where $U_{{\rm rad},j}$ is the total radiation energy density summed over all cells with distance $<r$ from the origin. Perfect agreement would consist of $E_{\rm num}(<r) = E_{\rm exact}(<r)$.

\begin{figure}[!t]
\centerline{\includegraphics[trim=0.75cm 5.5cm 2.25cm 6cm,clip,width=0.75\columnwidth]{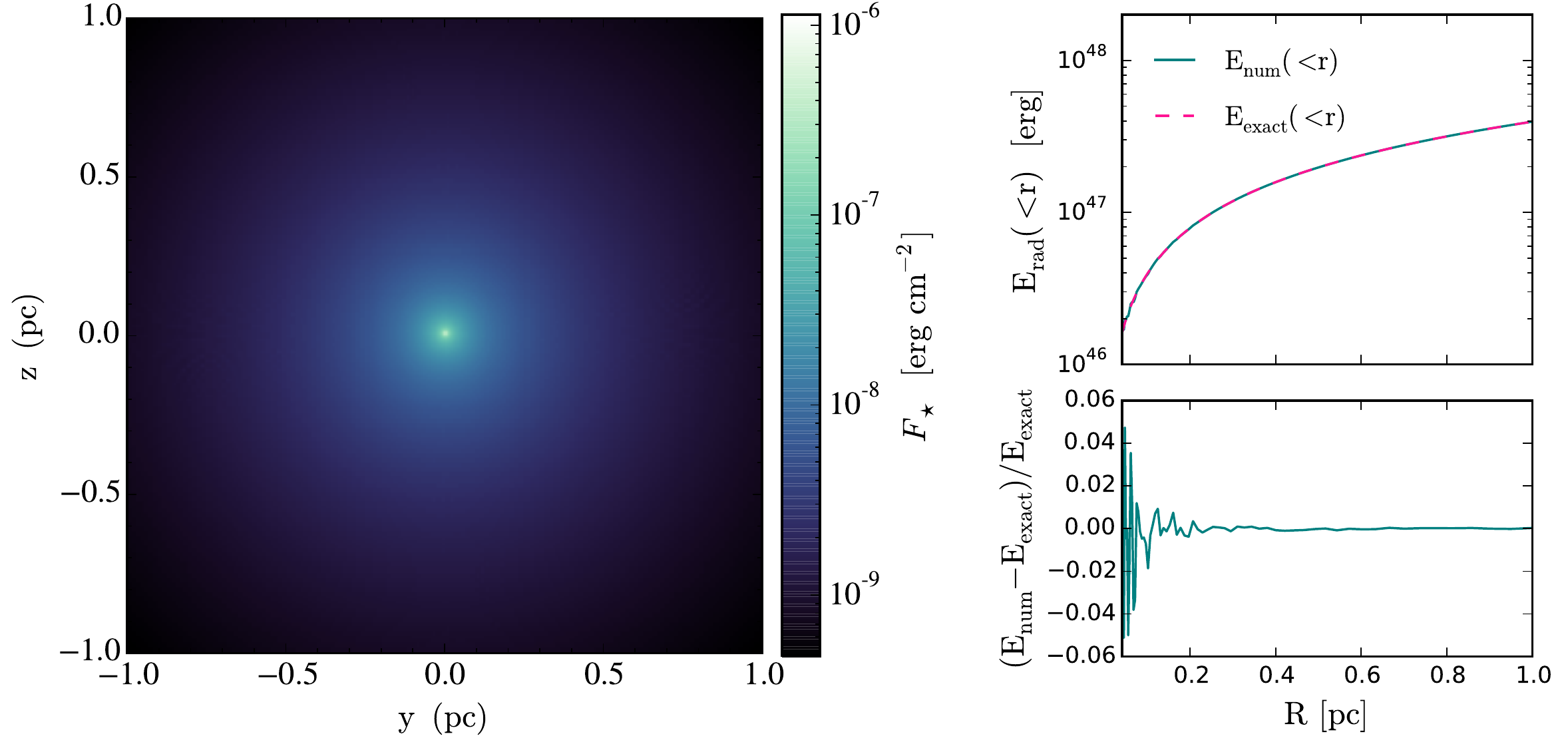}}
\caption{
\label{fig:eEnc}
Performance test for the adaptive ray trace. Left panel: Projection plot of the stellar radiation density  for a source with luminosity $10^6 \; L_{\rm \odot}$. The source flux falls off as $F(r) \propto r^{-2}$ as expected. Right panels: Comparison of the numerical and analytical results of the energy enclosed within radius $r$ (i.e, $E_{\rm num}(<r)$) for the same source. Top panel: The pink dashed line shows the exact analytical solution and the teal solid line is the numerical result from the adaptive ray trace. The bottom panel shows the residuals from the exact and numerical solutions.
}
\end{figure}

Our results are shown in Figure \ref{fig:eEnc}. The left panel shows the line-of-sight projected radiation energy density of the point source radiation field (i.e., Equation (\ref{eqn:ecell})) integrated over the line of sight, which drops off as $r^{-2}$ as expected.  The two right panels compare $E_{\rm num}(<r)$ and $E_{\rm exact}(< r)$. We find that the difference between the numerical and exact results is always $<5\%$, with the maximum error occurring close to the source where the resolution is poor. This error is expected because of the fact that we are using Cartesian rather than spherical grids.

\subsection{Radiation-Pressure-Dominated HII Region}
\label{sec:raddom}
Next we perform a test to illustrate the coupling of the radiative transfer from the adaptive ray trace with the hydrodynamics, based on a similarity solution obtained by \citet{krumholz2009b}. We consider an initially-uniform, cold gas with an isothermal equation of state. At time $t=0$ a point source of radiation with luminosity $L_{\rm \star}$ turns on and begins depositing momentum in the gas. We consider material that has a very high opacity to photons coming directly from the point source, but a very low opacity to any re-emitted photons. A real-world example of this would be interstellar dust absorbing ultraviolet photons from a star, and then re-emitting them as infrared light, to which the dust is essentially transparent.

Because the opacity is high, all of the radiation from the point source is absorbed in an extremely thin layer, but then escapes immediately. Thus the point source deposits radial momentum into the gas at a rate $dp/dt = L_{\rm \star}/c$. After a short time the material around the point source will have been swept into a thin shell of radius $r_{\rm sh}$ and mass $M_{\rm sh}=4\pi r_{\rm sh}^3 \rho_0/3$, where $\rho_0$ is the initial density. The shell obeys an equation of motion
\begin{equation}
\frac{d}{dt} \left( M_{\rm sh} \dot{r}_{\rm sh} \right) = \frac{L_{\rm \star}}{c}.
\end{equation}
This equation admits a similarly solution given by
\begin{equation}
\label{eqn:rsh1}
r_{\rm sh}(t) = 1.15 \left( \frac{n_0}{10^5 \, \rm{cm^{-3}}}\right)^{-1/4} \left(\frac{L_{\rm \star}}{10^6 \, L_{\rm \odot}}\right)^{1/4} \left( \frac{t}{\rm{Myr}}\right)^{1/2} \, \rm{pc} 
\end{equation}
\noindent
where $n_0=\rho_0/\mu m_{\rm p}$ is the number density, $\mu$ is the mean molecular weight which we set to 2.33 for molecular hydrogen and helium mixed in the usual cosmic ratio, and $m_{\rm p}$ is the proton mass.

To test the ability of our code to reproduce this solution, we consider a domain with a width of 1 pc, a  uniform number density of $n_0 = 10^{5} \; \rm{cm^{-3}}$ ($\rho_0= 3.89 \times 10^{-19}\; \rm{g \, cm^{-3}}$), and a point source of luminosity $L_{\star}=10^6 L_{\rm \odot}$ at the origin. We set the specific opacity to $\kappa/\rho = 10^6 \; \rm{cm^2 \;g^{-1}}$. We perform 3 tests  on non-adaptive grids with varying resolution ($64^3$, $128^3$, and $256^3$) to explore how the accuracy of the adaptive ray trace depends on resolution. 

\begin{figure}[!t]
\centerline{\includegraphics[trim=0.0cm 2.5cm 0.0cm 5.0cm,clip,width=0.75\textwidth]{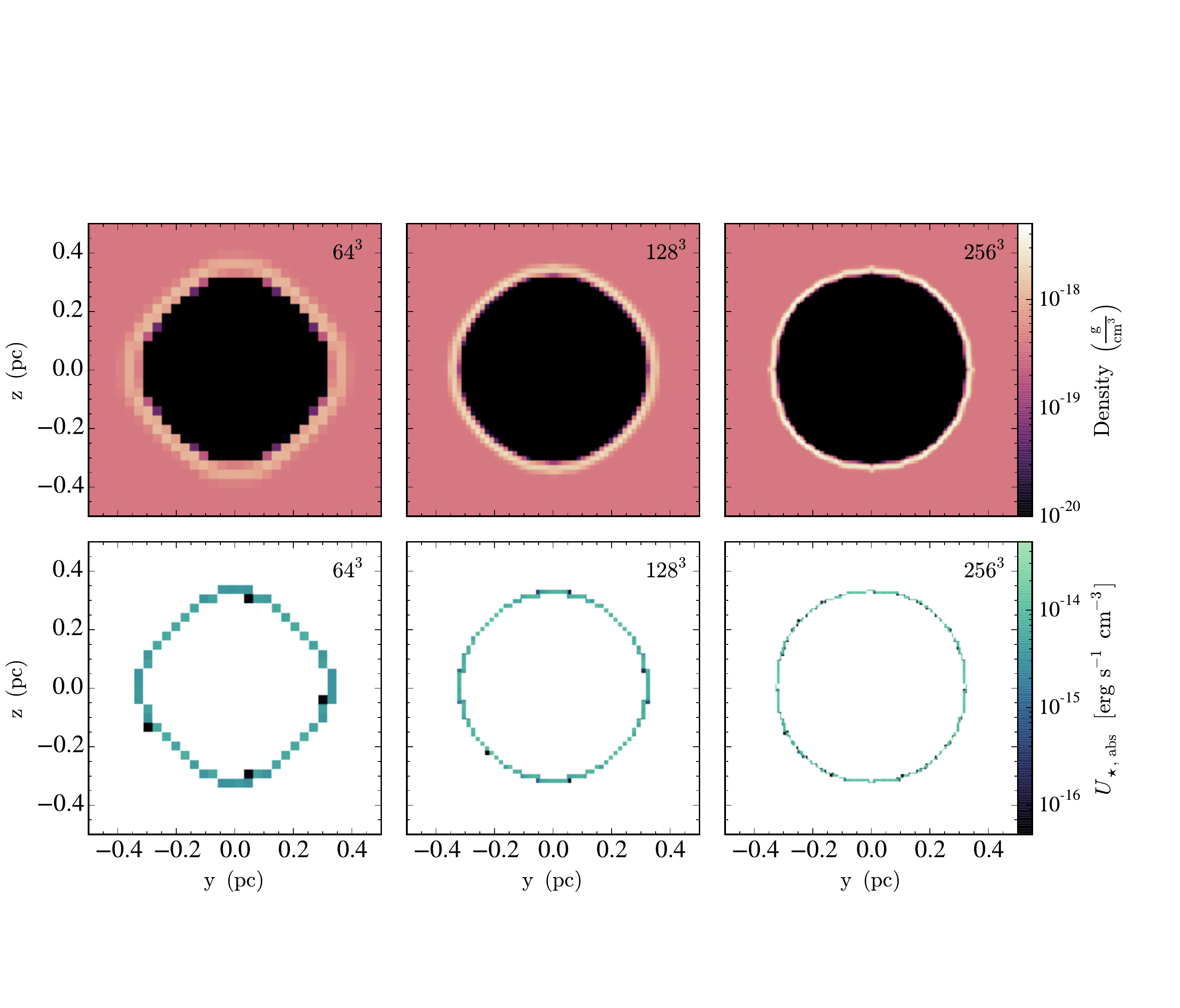}}
\caption{
\label{fig:raddom}
Results from the radiation-pressure-dominated \hii\ region test to demonstrate the performance of the adaptive ray trace coupled to the hydrodynamics. Top (bottom) panels show slice plots of the gas density (rate of absorbed radiation energy density per unit time) for our radiation dominated sphere test at three different uni-grid resolutions ($64^3$, $128^3$, and $256^3$) taken at $t=0.1$ Myr. As the bottom panels show, the direct radiation is absorbed only by the dense shell due to the high specific opacity, $\kappa=10^6 \; \rm{cm^2/g}$, used. 
}
\end{figure}

Figure \ref{fig:raddom} shows a snapshot of the simulation results at $t=0.1 \, \textrm{Myr}$. The top panels show the density slices and the bottom panels show the deposition rate of the stellar radiation energy density. We run the simulations to $t=0.35$ Myr, but at later times we develop the carbuncle instability which distorts the shape of shock waves that move along grid directions \citep{pandolfi2001,stone2008a}. One can eliminate this instability by implementing extra dissipation in grid-aligned flows \citep{stone2008a}, but since real applications are never perfectly grid-aligned, we have not done so here.

\begin{figure}[!t]
\centerline{\includegraphics[trim=1.0cm 0.0cm 1.0cm 0.5cm,clip,width=0.75\textwidth]{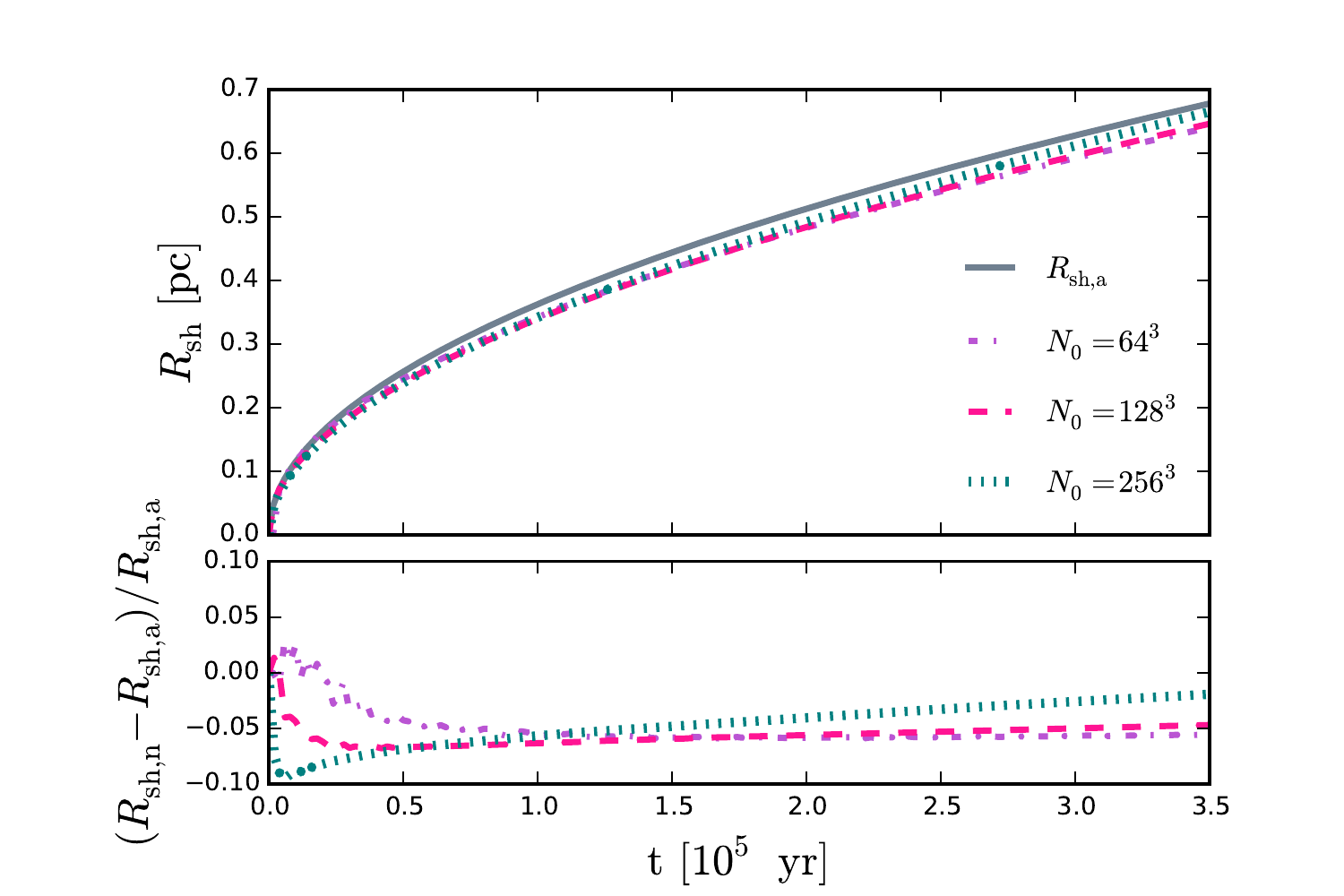}}
\caption{
\label{fig:rsh}
Shell position (top panel) and residuals (bottom panel) as compared to the analytical solution (i.e., eqn \ref{eqn:rsh1}) for our radiation dominated sphere test at three different uni-grid resolutions ($64^3$, $128^3$, and $256^3$). The largest deviations from the analytical solution occur at early times when the shell is located close to the source but the numerical result follows the analytical solution better as the shell expands.
}
\end{figure}

In Figure \ref{fig:rsh} we show how the radius of the shell in our simulation compares to the analytic similarity solution. We define the radius of the shell to be the density weighted average distance from the origin for cells where the density exceeds 1.5 $\rho_0$:
\begin{equation}
\label{eqn:rsh}
R_{\rm sh} = \frac{\sum_{{\rm cells},\, \rho_j > 1.5 \rho_{0}} \rho_j r_j}{\sum_{{\rm cells},\, \rho_j > 1.5 \rho_{0}} \rho_j}.
\end{equation}
The top panel shows the shell radius as a function of time for each resolution and for the analytic solution, while the bottom panel shows the residuals. As in Figure \ref{fig:eEnc}, the residuals are largest at early times when the shell is poorly resolved, but the agreement becomes excellent at later times. The accuracy of the solution also improves with increasing resolution, as expected.

%\subsection{\red{Hybrid Radiation Test}}
%Perform shadowing test similar to \citet{klassen2014a} in which we have two sources near a dense clump. Perform 4 tests with different physics:
%\begin{enumerate}
%\item no hydro, ray tracing only
%\item no hydro, ray tracing and FLD
%\item hydro, ray tracing only
%\item hydro, ray tracing and FLD
%\end{enumerate}

%Test Parameters:
%\begin{enumerate}
%\item Domain size
%\item Base resolution: $256^3$
%\item Refine cloud to level 1
%\item 0.1 pc 
%\item Have stars be 10 Msun, Lzams= 2.135255328529133e+37 erg/s, Rzams=2.743487473588309e11 cm, 
%\end{enumerate}

\section{Performance Tests}
\label{sec:perform}
It is important for our code to scale well with number of processors, especially for large simulations. Scaling tests demonstrate the efficiency of a parallel application when increasing the number of processors. In this section we present both weak and strong scaling tests to demonstrate the parallel performance of our adaptive ray trace algorithm. We also perform a strong scaling test for our hybrid radiation algorithm in section \ref{sec:msf} in an AMR simulation to demonstrate the scaling capability of \harm\ for a demanding, research application. For all tests, we set $\Phi_{\rm c}=4$ and also have the initial ray level set to 4 so that 3072 rays are initialized at the beginning of each ray trace step. All of the following performance tests were run on the Sandy Bridge nodes on the NASA NAS machine Pleiades.

\subsection{Weak Scaling}
\label{sec:ws}
Weak scaling tests demonstrate how well a parallel code scales with the number of processors while the workload assigned to each processor remains the same. For this purpose, we perform a weak scaling test on non-adaptive grids where each processor has one $32^3$ grid and one radiating sink particle. Each $32^3$ grid represents a (1 pc)$^3$ domain with constant gas density, $n=10^4 \, \rm cm^{-3}$, with a radiating source placed at the center. The physics modules we include are the adaptive ray trace and hydrodynamics. We set the opacity in all cells to zero so that no absorption of the radiation field occurs. To ensure that each processor performs the same amount of work with the ray trace, including the propagation of rays and the subsequent communication of rays to other processors, we terminate rays once they have traveled 0.6 pc from their originating source. This allows for rays to propagate to their neighboring grids and also enforces that all grids, except the grids along the domain edges, communicate the same number of rays to their neighbors. In short, the rays interact with the cells they cross but do not add energy or momenta to the fluid. 

\begin{figure}[!t]
\centerline{\includegraphics[trim=0.25cm 0.25cm 0.25cm 0.1cm,clip,width=0.75\textwidth]{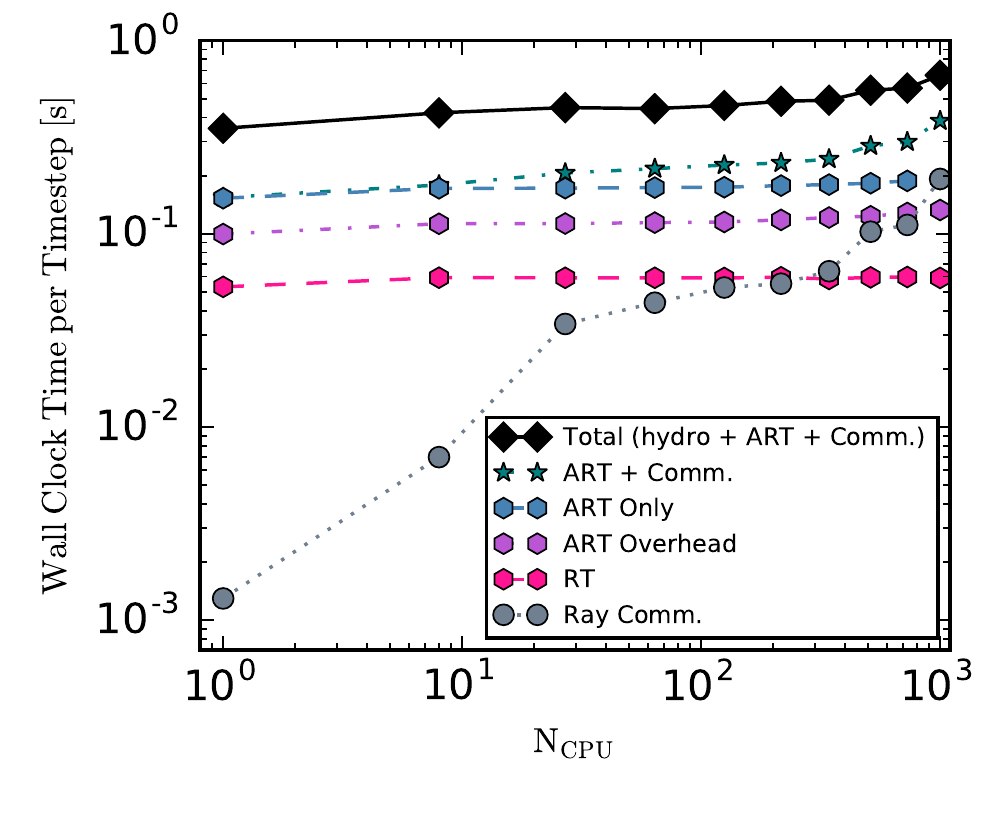}}
\caption{
\label{fig:ws}
Weak scaling test with one $32^3$ block per process. Each block is (1 pc)$^3$ and contains one radiating source at its center. Rays are terminated after they have traveled 0.6 pc from the source to ensure communication of rays to neighboring grids. Weak scaling results are shown for the ray communication (gray dotted line), ray tracing across cells (pink dashed line), overhead associated with the adaptive ray trace (purple dot-dashed line), adaptive ray trace excluding the communication of rays (blue line), the total cost of the adaptive ray trace including parallel communication of rays (teal dot-dashed line), and the total time spent on the hydrodynamics and adaptive ray tracing (solid black line). Ray communication is cheaper than the adaptive ray trace (ray trace and associated overhead) for $N_{\rm CPU} \lesssim 1000$ processors. The communication shows a $N^{0.67}_{\rm CPU}$ dependence.
}
\end{figure}

Our weak scaling tests were run on $N_{\rm CPU}=n^{3}$ processors, where n=[1,2,...,9,10], for 50 time steps per weak scaling test. The weak scaling test results are presented in Figure \ref{fig:ws} and show the total time spent per time step (black solid line) and the timing of the adaptive ray trace components: ray communication (gray dotted line), ray trace across cells (pink dashed line), adaptive ray trace overhead (i.e., locating ray grids, ray splitting, etc. -- purple dot-dashed line), total adaptive ray trace (ray trace and associated overhead -- blue dashed line), and the full adaptive ray trace which includes the ray communication (dot-dashed teal line). We note that a horizontal line denotes perfect weak scaling.

Our timing results show that the tracing of rays across cells has near-perfect weak scaling for all processor counts and that the adaptive ray trace overhead exhibits near perfect weak scaling until $\sim$216 processors.  We also find that the costs associated with the overhead are more expensive than ray tracing alone. Finally, our ray communication algorithm is cheaper than the ray tracing up to $\sim$343 processors and cheaper than the costs associated with the adaptive ray trace overhead up to $\sim$729 processors. The ray communication only becomes as expensive as the adaptive ray trace at $\sim$1000 processors. This is because our asynchronous communication algorithm, described in section \ref{sec:parallel}, has a $N_{\rm CPU}^{0.67}$ dependence. These results confirm that our communication algorithm is much more scalable and efficient when compared to previous methods. For example, the ray communication timing in \citet{wise2011a} followed a $N_{\rm CPU}^{1.5}$ dependence and became the dominant cost of the ray trace at only $\sim 200$ CPUs, despite the fact that their weak scaling test uses $64^3$ rather than $32^3$ blocks, and thus is significantly less stringent than ours.

\subsection{Strong Scaling}
\label{sec:ss}

Strong scaling demonstrates how well the code performs as the number of processors for a given problem increases while the total workload remains the same. We perform three tests to demonstrate the strong scalability of the adaptive ray trace and the \harm\ algorithm in a demanding, research application. The first two tests are performed on non-adaptive grids. The first test measures the strong scalability of the adaptive ray trace based on the number of cells that interact with rays (see section \ref{sec:ss1}) while the second test focuses on the strong scalability of the adaptive ray trace as a function of the number of frequency bins used (see section \ref{sec:ss2}). The third test shows the parallel performance of \harm\ in a demanding, research AMR simulation. In addition to \harm, this test also includes other physics modules in \orion\ such as hydrodynamics, self-gravity, and sub-grid star particles. Each test and their results are described below.

\subsubsection{Uni-grid Ray Trace Test with Varying Termination Lengths}
\label{sec:ss1}
We first use a setup similar to a single instance of our weak scaling test: a (1 pc)$^3$ domain with a single point source placed at its center. The resolution of the computational domain is $256^3$ cells and each grid consists of $16^3$ cells yielding a total of 4096 grids, with no adaptivity. We perform three sub-tests with this setup in which the rays are destroyed after traveling 0.2 pc from their source, 0.4 pc from their source, or allowed to trasverse the entire domain, respectively. These calculations were performed on $N_{\rm CPU}=2^n$ processors with n=[2,...,9,10] for 5 time steps per test. 

\begin{figure}[!t]
\centerline{\includegraphics[trim=0.25cm 0.25cm 0.25cm 0cm,clip,width=0.75\textwidth]{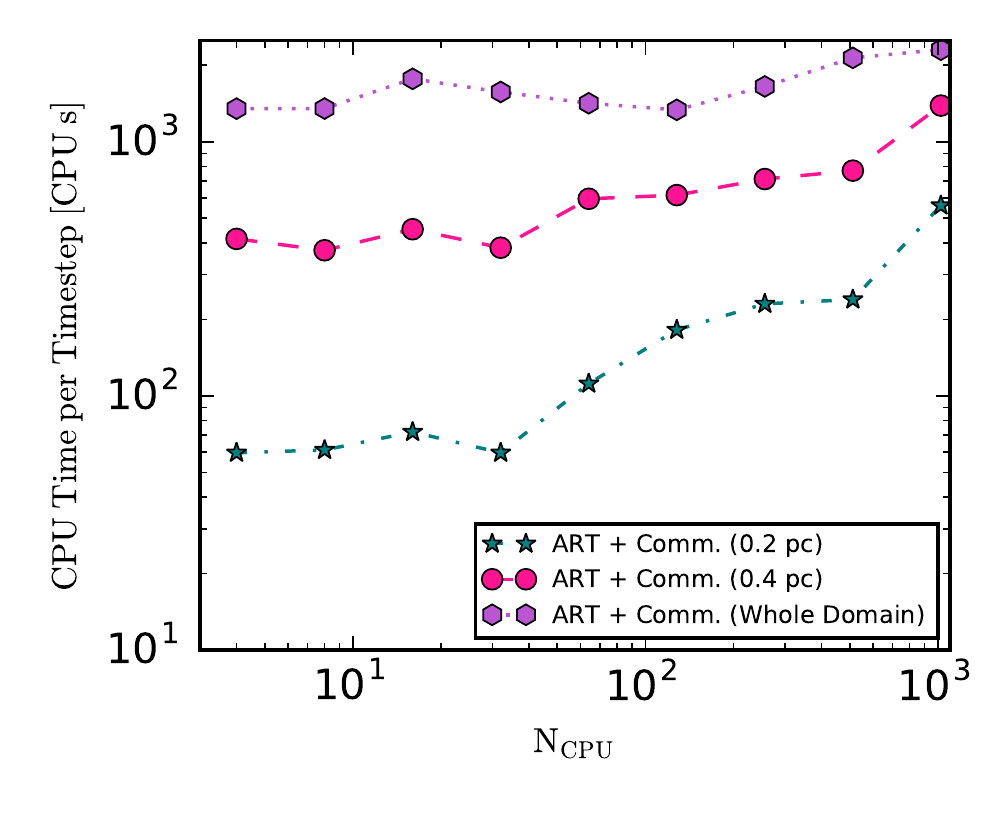}}
\caption{
\label{fig:ss1}
Strong scaling test with a $256^3$ uni-grid calculation with a radiating point source at the center. We performed two tests where the rays are terminated after they travel 0.2 or 0.4 pc from their source and a third test where the rays trasverse the entire domain. Perfect strong scaling would yield a flat line for each test. Our results show that the strong scaling performance improves as the volume that the rays interact with increases and that near-perfect strong scaling is attained when the rays trasverse the entire domain. 
}
\end{figure}

Our strong scaling results are presented in Figure \ref{fig:ss1}, which shows the total CPU time per time step, $t_{\rm CPU}$, for the adaptive ray trace. (Note that, whereas in Figure \ref{fig:ws} we plotted the time per processor, here we plot the total time summed over all processors, so that perfect scaling would again appear as a flat horizontal line.) To better quantify the results, we perform a $\chi^2$ fit of our measured results to the functional dependence $t_{\rm CPU} \propto N_{\rm CPU}^a$; perfect strong scaling would be $a=0$. We report these results in Table \ref{tab:ss1}.

When we allow rays to traverse the entire computational domain, we find near-perfect strong scaling out to 1024 processors: $t_{\rm CPU} \propto N_{\rm CPU}^{0.084}$. As we lower the distance that rays propagate, the scaling deteriorates, for the obvious reason that processors which are assigned computational domains that rays do not reach are simply idle because they do not contribute to the ray trace computation. Indeed, we also report the fraction of the computational volume over which rays propagate in Table \ref{tab:ss1}, and it is clear that the scaling is worse when this value is small.

\begin{table*}
	\begin{center}
	\caption{
	\label{tab:ss1}
Fitted scaling results from our uni-grid strong scaling tests presented in Figure \ref{fig:ss1}, together with the fraction of the computational volume over which the ray trace is performed.  A value of $a=0$ would imply perfect strong scaling.
}
	\begin{tabular}{ l c  c }
	\hline	
	$D_{\rm Ray}$ & Ray-interaction Volume & $a \; (N^a_{\rm CPU})$\\
	\hline
	\hline
	0.2 pc & 0.0335 & 0.52 \\
	0.4 pc & 0.2681 & 0.27\\
	Whole & 1 & 0.084\\
	\hline
	\end{tabular}
	\end{center}
\end{table*}

\subsubsection{Timing with Varying Frequency Bins}
\label{sec:ss2}
Our adaptive ray trace algorithm allows for an arbitrary number of frequency bins $N_{\rm \nu}$. Each ray has two arrays that contain $N_{\rm \nu}$ doubles (size 8 bytes) that hold the ray's initial and current frequency-dependent luminosities, respectively. The choice of $N_\nu$ impacts the cost of the computation in two ways: (1) the ray trace operations must loop over all frequency bins when creating rays, advancing them across cells, and checking if they become extinct due to absorption by the fluid; and (2) MPI communication operations depend on the size of the message that is being sent and/or received. Therefore, increasing the number of frequency bins for the adaptive ray trace will lead to an increased cost in the overhead associated with the advancement and communication of rays. 

To test the scaling efficiency of the adaptive ray trace as a function of $N_{\rm \nu}$ we ran a series of tests where we vary the number of frequency bins. Our initial setup of our test problem is the same as the strong scaling test discussed in section \ref{sec:ss1} in which a radiating source is at the center of a (1 pc)$^3$ box.  We terminate the rays after they have travelled 0.5 pc and perform tests for $N_{\rm \nu} = (1, 2, 8, 16, 20, 32, 48, 64)$ frequency bins. Our base grid is $256^3$ and we ran our scaling tests on 128 processors for 50 time steps per test. Perfect strong scaling on this test would be a computational cost proportional to $N_{\rm \nu}$, since the number of ray-cell interactions is linear in $N_{\rm \nu}$.

\begin{figure}[!t]
\centerline{\includegraphics[trim=0.25cm 0.0cm 0.0cm 0.0cm,clip,width=0.75\textwidth]{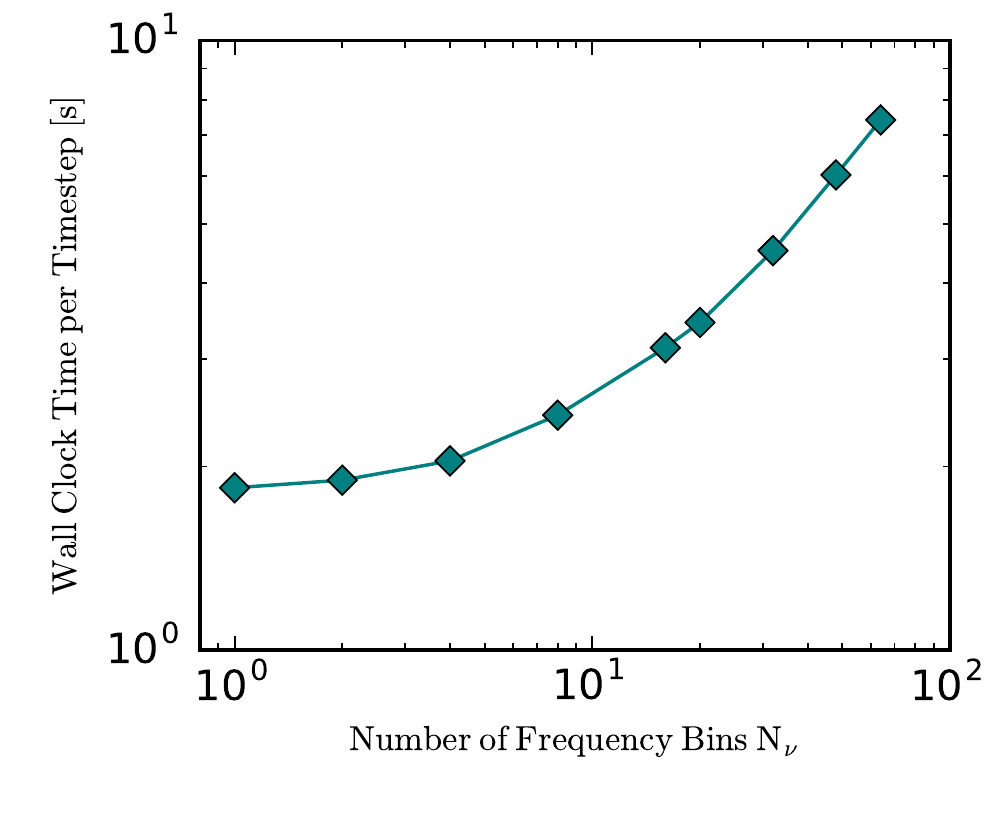}}
\caption{
\label{fig:fbs}
Frequency bin scaling test where we have varied the number of frequency bins $N_{\rm \nu}$. There is one source at the center of a $(128)^3$ domain and we truncate rays once they have traveled 0.5 pc. The cost of the ray trace rises with $N_{\rm \nu}$, as expected, but this effect is small for low $N_{\rm \nu}$.
}
\end{figure}

Our strong scaling results with varying $N_{\rm \nu}$ are shown in Figure \ref{fig:fbs}. We find that the wall clock time spent per ray trace increases with $N_\nu$ as expected, but that this increase is highly-sublinear, particularly at small $N_{\rm \nu}$.  We find $t_{\rm \nu} \propto N_{\rm \nu}^{0.14}$ for 1-8 frequency bins and $t_{\rm \nu} \propto N_{\rm \nu}^{0.65}$ for 16-64 frequency bins. While these results might at first seem surprising, they make sense when we recall that the overhead associated with the ray tracing -- solving the geometric problem of finding the paths of rays through cells and grids, the probing and handshaking parts of the communication steps -- does not scale with $N_{\rm \nu}$. As we increase $N_{\rm \nu}$, this overhead is ``amortized" over a larger number of frequency bins, and thus we obtain what appears to be better-than-perfect strong scaling. As the number of frequency bins increases, this effect becomes less important, and the parts of the computation that do scale with $N_{\rm \nu}$ -- computing the opacities of cells and updating fluxes, transferring flux data between processors -- begin to dominate. For sufficiently large $N_{\rm \nu}$ we do begin to approach the expected $N_\nu^{1.0}$ scaling, but our results thus far demonstrate that we can use up to $\sim 10$ frequency bins at near-zero additional cost, and several tens at only modest cost, compared to the single-frequency case.

\subsubsection{AMR Simulation: Application to High-Mass Star Formation}
\label{sec:msf}
Our final strong scaling test is an AMR simulation that includes hydrodynamics, self-gravity, radiative transfer, and radiating sink particles to demonstrate how our new \harm\ algorithm scales in a demanding research application. Here we perform strong scaling tests for two different outputs from an \orion\ AMR simulation of the formation of a high-mass stellar system. The results of this simulation will be presented in \citet{rosen2016b} but we briefly summarize our problem setup here. 

Our initial condition is a rotating, laminar 150 $M_{\rm \odot}$ molecular core with radius 0.1 pc. The core follows a $\rho(r) \propto r^{-3/2}$ density profile. We use a domain size of 0.4 pc on each side, a base resolution of $128^3$ and five levels of refinement which yields a maximum resolution of 20 AU on the finest level. To properly model the absorption of the direct radiation field from stars we use the frequency dependent stellar atmosphere profiles from \citet{lejeune1997a} to model the stellar spectra. Our choice of the opacities depend on whether the primary absorber is dust or molecular gas. Dust is the primary absorber for gas temperatures below $T_{\rm sub} = 1500$ K (i.e., the temperature at which dust sublimes) \citep{semenov2003a} while molecular hydrogen is the primary absorber for gas temperatures within $T_{\rm sub} \le T < T_{\rm \hii}$ where $T_{\rm \hii} \approx 10^4$ K is the temperature at which we expect hydrogen to become fully ionized, and thus to have the usual Thompson opacity for electron scattering. If the primary absorber is dust we use the frequency dependent dust opacities from \citet{weingartner2001a} (their $R_{\rm v}=5.5$ extinction curve), if it is molecular hydrogen we set the molecular gas opacity to 0.01 $\rm{cm^2 \, g^{-1}}$, and if $T \ge T_{\rm \hii}$ we set the opacity to zero. The last of these is a numerical convenience, because we have not implemented scattering or ionization chemistry, and because the regions in our computation with $T>T_{\rm \hii}$ generally contain so little mass they will be optically thin anyway. We assume a dust-to-gas ratio of 0.01 and choose $N_{\rm \nu}=10$.

\begin{figure}[!t]
\centerline{\includegraphics[trim=1cm 0.0cm 1cm 0.2cm,clip,width=0.75\textwidth]{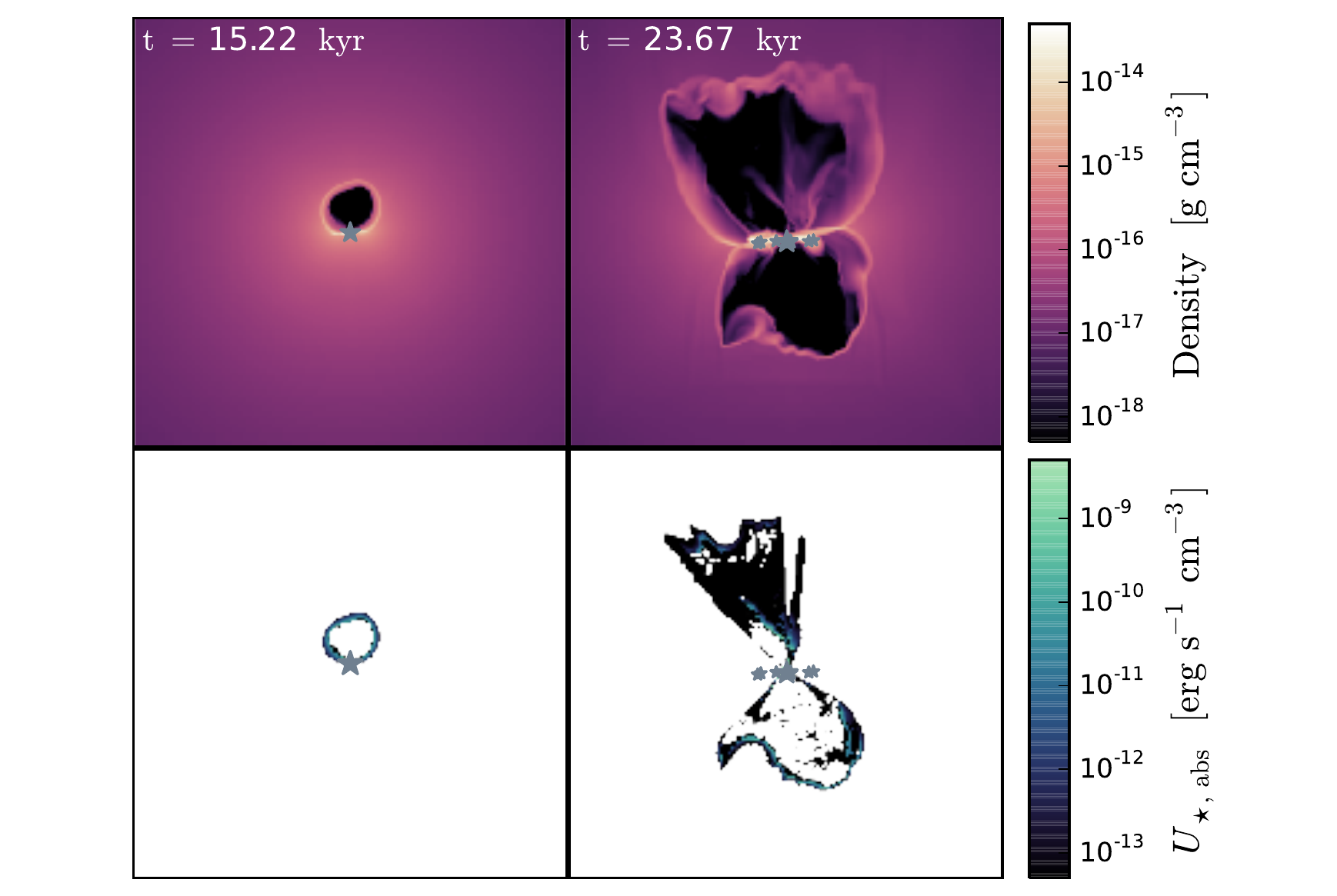}}
\caption{
\label{fig:HM}
Example AMR simulation that uses our \harm\ algorithm. Here we show slice plots along the x-direction of the mass density (top) and absorbed direct radiation energy density (bottom) for two snapshots of a simulation of the formation a high mass star system. Gray stars denote the location of the stars, with the most massive star being largest. The left (right) panels show the snapshot when the simulation has progressed to 15.22 kyrs (23.67 kyrs) where the most massive star is 16.59 $M_{\rm \odot}$ (33.57 $M_{\rm \odot}$).
}
\end{figure}

At $t=0$ the molecular core begins to gravitationally collapse. As the core collapses a star forms at the center and continues to grow in mass via accretion. An accretion disk forms around the star due to conservation of angular momentum of the infalling material. Gravitational instabilities develop in the disk causing it to fragment into companion stars. The absorption of energy and momenta from the direct stellar radiation field and the diffuse dust-reprocessed radiation field from the fluid results in low-density, radiation pressure dominated bubbles near the poles of the most massive star that expand with time. Figure \ref{fig:HM} shows slices parallel to the x-direction of the gas density (top panels) and absorbed direct radiation energy density (bottom panels) by the dust and gas for two different snapshots of this simulation at $t=15.22$ kyrs  and $t=23.67$ kyrs. We only show the central (8000 AU)$^{2}$ region of the computational domain because the majority of the domain is not affected by the direct radiation field. The most massive stars in these snapshots are 16.59 $M_{\rm \odot}$ and 33.57 $M_{\rm \odot}$, respectively. The early snapshot contains one star while the later snapshot contains eight stars where the companions range from $0.01-1.48$ $M_{\rm \odot}$ in stellar mass. These snapshots represent typical ``early" and ``late" stages of the computation, with the latter being much more computationally expensive due to the larger number of sources and the greater distances that the direct radiation can propagate before being absorbed. We note that both the early and late stages are strong tests of the scalability, because the radiating sources are confined to a small portion of the computational volume, rather than being scattered throughout (c.f., the test presented in \citet{wise2011a}, which used a cosmological simulation where point sources were distributed nearly-isotropically.)

\begin{figure}[!t]
\centerline{\includegraphics[trim=0.5cm 0.7cm 0.25cm 0.5cm,clip,width=0.5\textwidth]{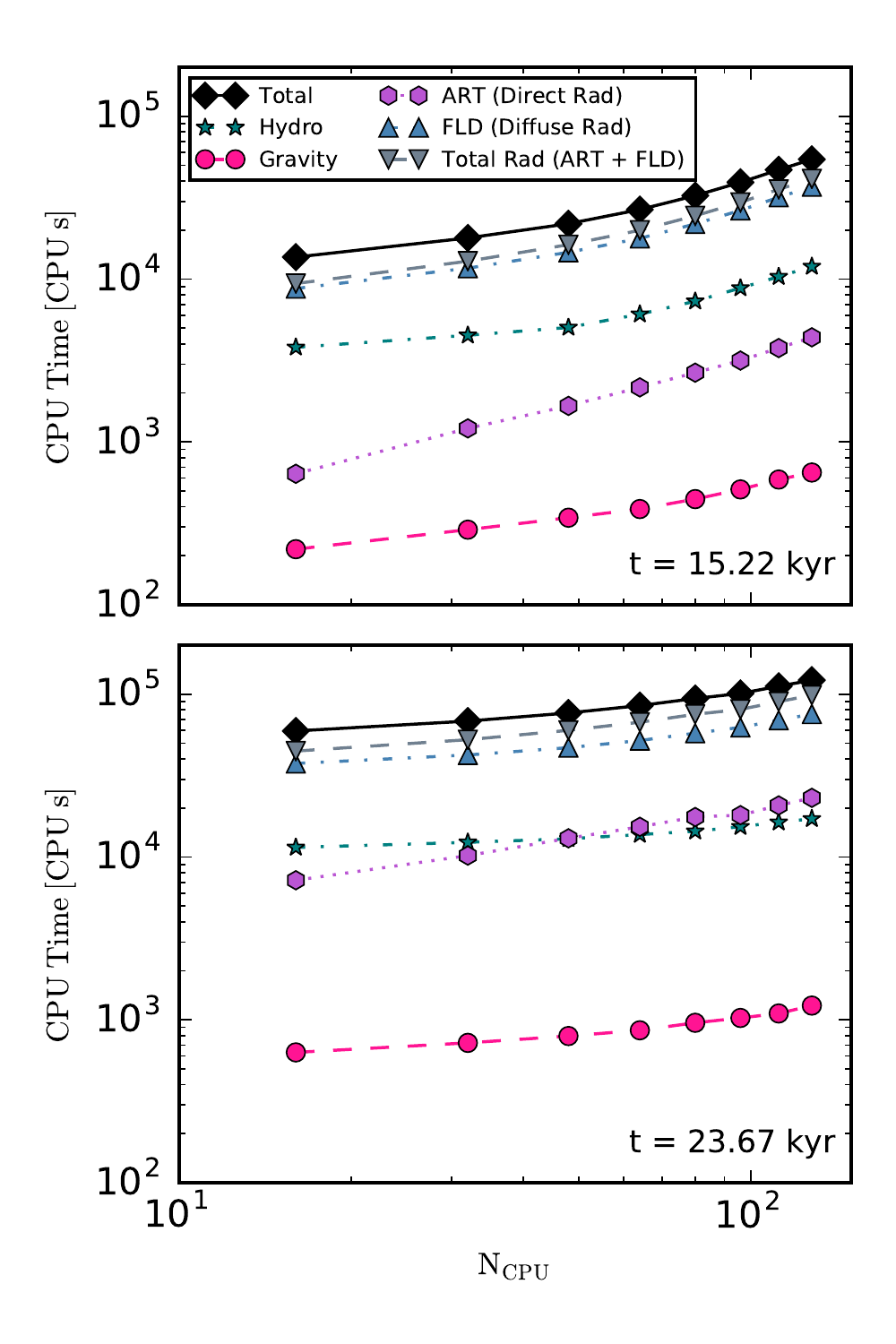}}
\caption{
\label{fig:ssHM}
Strong scaling test with a $128^3$ AMR simulation with 5 levels of refinement of the formation of a massive star system shown at two different simulation outputs from 15.22 kyrs (top) and 23.67 kyrs (bottom). The early (late) snapshot has 448 (1137) grids. The bottom panel shows that the scalability of the adaptive ray trace increases as the simulation progresses because rays interact with a larger volume of the computational domain (e.g., see Figure \ref{fig:HM}).
%\red{Notes: change labels on bottom axis, make panels in two columns instead of rows}
}
\end{figure}

Our strong scaling results are shown in Figure \ref{fig:ssHM}, where we measure the time spent on the hydrodynamics, gravity, FLD, adaptive ray trace, and the total radiation module (adaptive ray trace and FLD). A horizontal line would correspond to perfect  strong scaling. We ran each timing test for five time steps on $N_{\rm CPU} = 16 \times n$ processors where $n=[1,2,...,8]$. The early snapshot contains 448 grids and the later snapshot contains 1137 grids at the beginning of each test. The top panel shows the timing results for the early snapshot and the bottom panel shows the results for the later snapshot. Comparison of the two panels show that the scalability for all modules in \orion\ become better at later times, especially for the adaptive ray trace when more grids are processing rays. This is due to the increase in number of grids per processor which reduces the MPI communication costs. A general rule of thumb for patch-based AMR methods such as \orion\ is that the the code is efficient at $\sim 4$ grids per CPU or more, and our tests are consistent with this. We find that our timing results for the adaptive ray trace follows $t_{\rm WC, ART} \propto N_{\rm CPU}^{0.97}$ for the early snapshot and $t_{\rm WC, ART} \propto N_{\rm CPU}^{0.56}$ for the later snapshot. These results agree with our strong scaling results from section \ref{sec:ss1} which showed that our parallelization procedure for the adaptive ray trace becomes more efficient as the number of grids that interact with rays increases. We find that the moment method, FLD in our case, is the most expensive module while gravity is the cheapest, and that the adaptive ray trace can be cheaper and/or about the same expense as the hydrodynamics.

%Shape of line depends on how cores get distributed to CPUs. When region of ray trace is small - get bad scaling because most CPUs do not use the ray trace. 

%Emphasize: when rays go significant distance such that much of the computational domain works on the ray trace we get good strong scaling up to 512 processors this is as good a one can hope for.

\section{Summary}
\label{sec:sum}
In this paper, we have presented our implementation of \harm\ -- a new highly-parallel multi-frequency hybrid radiation hydrodynamics module that combines an adaptive long characteristics method for the (direct) radial radiation field from point sources with a moment method that handles the (thermal) diffuse radiation field produced by a volume-filling fluid. Our new method is designed to be used with adaptive grids and is not limited to specific geometries. We have coupled \harm\ to the hydrodynamics in the astrophysical AMR code \orion\ which includes flux limited diffusion, but our method can be applied to any AMR hydrodynamics code that has asynchronous time stepping and can incorporate any moment method. Although our implementation is not the first hybrid radiation scheme implemented in an AMR code, it is more accurate than previous methods because it uses long rather than hybrid characteristics. Furthermore, our new algorithm can be used in a variety of radiation hydrodynamics problems in which the radiation from point sources and diffuse radiation field from the fluid should be modelled. Such examples are the study of the formation of isolated high-mass stars and clustered star formation in the dusty interstellar medium. 

One of the major difficulties with incorporating a long characteristics method in an AMR code that allows for a general geometry, where the hydrodynamics is parallelized by domain decomposition, is the parallel communication of rays. This is because ray tracing is a highly serial process and each ray will usually cross multiple processor domains. In order to avoid significant communication overheads and serial bottlenecks that often occur with long characteristics methods we have implemented a new completely asynchronous and non-blocking communication algorithm for ray communication. We performed a variety of weak and strong scaling tests of this method, and found that its performance is dramatically improved compared to previous long characteristics methods. In idealized tests without adaptive grids we obtain near-perfect weak scaling out to $>1000$ cores, and, in problems where the characteristic trace covers the entire computational domain, near-perfect strong scaling as well. Previous implementations became communications-bound at processor counts a factor of $\sim 4$ smaller than this. In a realistic, demanding research application with a complex, adaptive grid geometry, and using 10 frequency bins for the characteristic trace, we find excellent scaling as long as there are at least $\sim 3-4$ grids per CPU, and we find that the cost of adaptive ray tracing is smaller than or comparable to hydrodynamics, and significantly cheaper than flux limited diffusion.

Since \harm\ works for adaptive grids in a general geometry, it can be used in a variety of high-resolution simulations that require radiative transfer. Our implementation in \orion\ will be made public in an upcoming release of the \orion\ code, and the \harm\ source code will be made available immediately upon request to any developers who are interested in implementing \harm\ in their own AMR codes.

\section*{Acknowledgements}
\noindent
We thank Nathan Roth, Andrew Myers, and Pak Shing Li for their assistance in integrating \harm\ into \orion, Chris McKee for helpful comments throughout the development process, and John Wise and Christian Baczynski for comments and advice. ALR and MRK acknowledge support from the National Aeronautics and Space Administration (NASA) through Hubble Archival Research grant HST-AR-13265.02-A issued by the Space Telescope Science Institute, which is operated by the Association of Universities for Research in Astronomy, Inc., under NASA contract NAS 5-26555 and Chandra Theory Grant Award Number TM5-16007X issued by the Chandra X-ray Observatory Center, which is operated by the Smithsonian Astrophysical Observatory for and on behalf of NASA under contract NAS8-03060. ALR and ATL acknowledge support from the NSF Graduate Research Fellowship Program. RIK acknowledges support from NASA through ATP grant NNX13AB84G, the NSF through grant AST-1211729 and the US Department of Energy at the Lawrence Livermore National Laboratory under contract DE-AC52-07NA27344. MRK and RIK acknowledge support from NASA TCAN grant NNX-14AB52G. MRK acknowledges support from Australian Research Council grant DP160100695.
\bibliographystyle{elsarticle-num-names}
\bibliography{numPaper_arxiv_ALR}
%\bibliography{../../RefsALR/refs}

\end{document}